\documentclass[11pt,a4paper]{scrartcl}

\usepackage{CLICdp}
\usepackage{dirtytalk}
\usepackage{textpos}
\usepackage{CLICdp_definitions}
\usepackage{textcomp}

\title{ Measurement of the H to ZZ branching fraction at 350 GeV and 3 TeV CLIC }

\clicdpconf{2021}{001}  

\date{\today}

\addauthor{N. Vuka\u{s}inovi\'{c}\thanks{nvukasinovic@vin.bg.ac.rs}}{\institute{1}}
\addauthor{I. Bo\v{z}ovi\'{c}-Jelisav\u{c}i\'{c}}{\institute{1}}
\addauthor{I. Smiljani\'{c}}{\institute{1}}
\addauthor{G. Ka\u{c}arevi\'{c}}{\institute{1}}
\addauthor{G. Milutinovi\'{c}-Dumbelovi\'{c}}{\institute{1}}
\addauthor{T. Agatonovi\'{c}-Jovin}{\institute{1}}
\addauthor{M. Radulovi\'{c}}{\institute{2}}
\addauthor{J. Stevanovi\'{c}}{\institute{2}}

\addinstitute{1}{\say{VIN\u{C}A} Institute of Nuclear Sciences - National Institute of the Republic of Serbia, University of Belgrade, Belgrade, Serbia}\break
\addinstitute{2}{Faculty of Science, University of Kragujevac, Kragujevac, Serbia}\break

\onbehalfof{the CLICdp Collaboration} 
 
\abstract{In this paper we present results of the determination of the statistical precision of the branching fraction
measurement, for Higgs decaying to ZZ$^\ast$ pairs at 3 TeV and 350 GeV CLIC. Measurements are simulated with
the CLIC\_ILD detector model, taking into consideration all relevant physics and beam-induced background
processes. It is shown that the product of the branching fraction BR(${H\rightarrow\thinspace ZZ^\ast}$) and the Higgs production cross-section can be measured with a relative statistical uncertainty of 3\% (18\%) at 3 TeV (350 GeV)
center-of-mass energy, using semileptonic final states and assuming an integrated luminosity of 5 (1) ab$^{-1}$.} 

\titlecomment{Talk presented at the International Workshop on Future Linear Colliders (LCWS2021), 15-18 March 2021.
C21-03-15.1.}

\graphicspath{ {./logos/}{./figures/} }

\begin{document}

\titlepage


\section{Introduction} 
\begin{figure}[h]
  \includegraphics[width=0.8\textwidth]{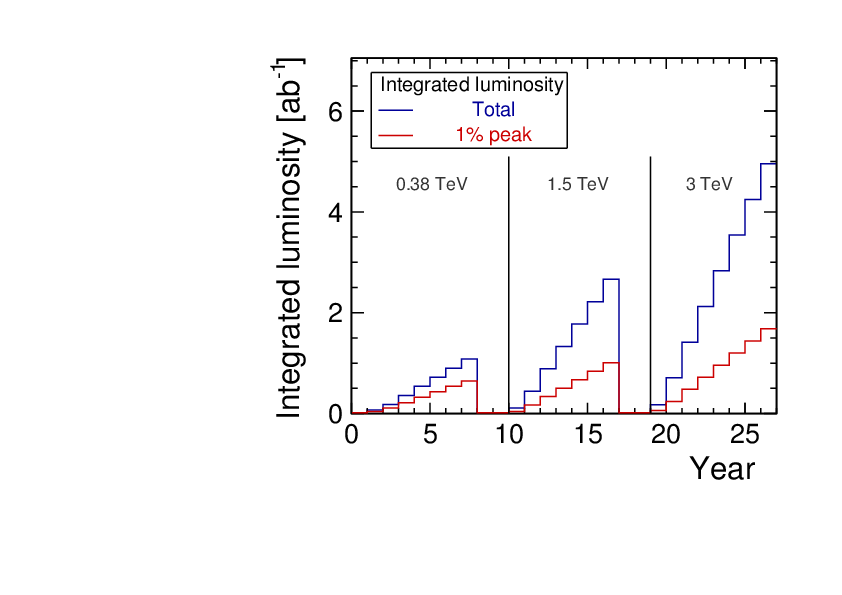}
   \vspace{-1cm}
  \caption{Luminosity per year in staged CLIC scenario. Due to the beamstrahlung the CLIC beam spectrum has a low-energy tail, so both the total luminosity per year and the luminosity collected above 99\% of the nominal $\sqrt{s}$ (labelled `1\% peak'), are shown. \label{fig:lumi} }
 \end{figure}

The Compact LInear Collider (CLIC) is a mature option for a future Higgs factory at CERN. If approved, CLIC could be ready for construction in 2026, with the first collisions in 2035 \cite{1}. CLIC is foreseen as a staged
machine that will run at center-of-mass energies: 380 GeV\footnote{For the first CLIC energy stage at 380 GeV, additional running time devoted to a t$\bar{\text{t}}$ threshold scan near 350 GeV is foreseen. Energy of 1.5 TeV
corresponds to the maximal center-of-mass energy reach with a single drive-beam complex and thus it is used in analyses as well as the initially proposed 1.4 TeV center-of-mass energy.}, 1.5 TeV\footnotemark[\value{footnote}] and 3 TeV with the corresponding
integrated luminosities of 1 ab$^{-1}$, 2.5 ab$^{-1}$ and 5 ab$^{-1}$, respectively (\cref{fig:lumi} ~\cite{2}).

\begin{figure}
  \centering
  \begin{subfigure}{0.48\textwidth}
    \includegraphics[width=\textwidth]{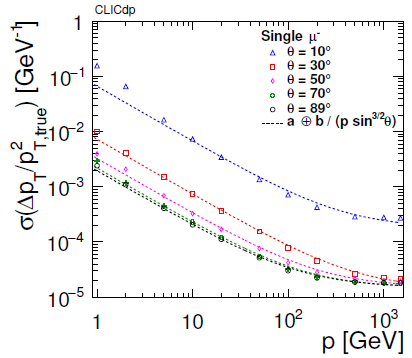}
    \caption{}
    \label{fig:pt}
  \end{subfigure}
  \hfill
  \begin{subfigure}{0.48\textwidth}
    \includegraphics[width=\textwidth]{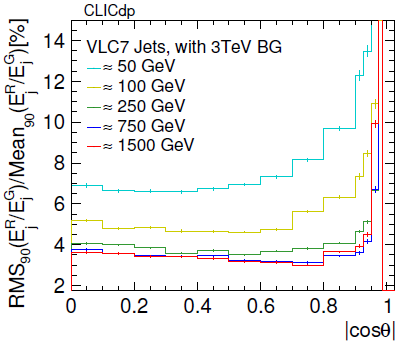}
    \caption{}
  \label{fig:cosj}
  \end{subfigure}
  \caption{Transverse momentum resolution as a function of momentum for muons, at various polar angles $\mathrm{\theta}$ \subref{fig:pt}. Jet
  energy resolution for different jet energies as a function of the polar angle ($|\text{cos}\theta|$) in the presence of overlaid $\mathrm{\gamma\gamma \rightarrow}$ hadron background. For the jet clustering, VLC algorithm ~\cite{9} with a jet cone radius R = 0.7 (VLC7) is used \subref{fig:cosj}.}\label{fig:2}
\end{figure}

The CLIC project combines a novel two-beam acceleration scheme, with a normal-conducting modular accelerator,
that has been demonstrated at the CTF3 CLIC test facility at CERN \cite{3}, along with the functionality of the main
accelerator components. The drive beam is a high-current beam (about 100 A) which generates a radio-frequency field
(12 GHz) that is transferred to the acceleration cavities of the main linac. In this way, conventional acceleration cavities achieve a high accelerating gradient of 100 MV/m. In the main linacs, the beam is accelerated from
190 GeV to 1.5 TeV energy. In order to maximize the reach of the CLIC physics programme, equal amounts of -80\%
and +80\% electron beam polarisation are foreseen at the initial energy stage. At higher-energy stages, a sharing of
the running time for -80\% and +80\% electron-beam polarisation is optimized in the ratio of 80:20 ~\cite{4}.
A detector for CLIC is being developed based on a broad range of full-simulation and experimental studies. In its latest
model (CLICdet ~\cite{5}) it comprises all-silicon vertexing and tracking components, compact Electromagnetic (ECAL)
and Hadronic (HCAL) calorimeters, all placed within a magnetic field of 4 T. The excellent performance of the
tracking system enables measurement of transverse momenta (p$_{\mathrm{T}}$) with a resolution $\mathrm{\sigma_{p_{T}}/{p_{T}^{2}}}$ of up to 2 $\cdot$ 10$^{-5}$ GeV$^{-1}$ for the high-energy charged particles in the central (barrel) detector region (\cref{fig:pt}, ~\cite{6}). Highly-granular calorimeters enable implementation of a Particle Flow Algorithm (PFA) ~\cite{7} allowing separation of jets that originate from Higgs and vector bosons Z$^{0}$ and W$^{\pm}$. For jet energies of 50 GeV the jet energy resolution is about 5\%, while for jet energies above 100 GeV jet-energy resolution is better than 3.5\% as illustrated in \cref{fig:cosj} ~\cite{6}. The CLIC\_ILD detector model ~\cite{8}
meets similar performances versus lepton identification efficiency and jet-energy reconstruction of relevance for this
study. 

\section{Higgs to ZZ measurements at CLIC}

\begin{figure}
  \centering 
  \includegraphics[width=0.48\textwidth]{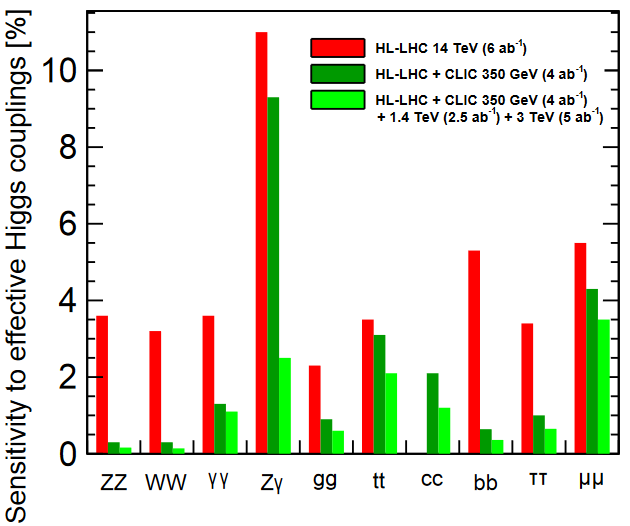}
  \caption{Projections of the model-dependent measurement of the Higgs couplings with a longer first stage CLIC, combined with the projected HL-LHC sensitivities. \label{fig:clic_eft} }
\end{figure}

Running at three different center-of-mass energies, CLIC can exploit several Higgs production
mechanisms. This approach contributes to larger statistics of produced Higgs bosons, including double-
Higgs production, and consequently to more precise determination of the Higgs boson properties (Higgs
mass, width, couplings, etc.). At 350 (380) GeV Higgsstrahlung is the dominant Higgs production
mechanism, while at higher energies WW-fusion takes over. Individual ${\mathrm{\sigma \times BR}}$ measurements at all energy
stages serve as input to global fits, either model-independent or model-dependent, to extract the Higgs
couplings with the utmost precision. Already the fit of data to be collected in the first energy stage of CLIC
operation gives better precision than HL-LHC, in particular for the Higgs couplings to c, b, W and Z (\cref{fig:clic_eft}, ~\cite{10}).
From \cref{fig:clic_eft} one also reads that the Higgs to ZZ couplings can be measured with a statistical uncertainty
of several permille.

\section{H$\mathrm{\rightarrow ZZ^\ast}$ analyses $@$ 350 GeV and 3 TeV}

\begin{figure}
  \centering
  \begin{subfigure}[h!]{0.38\textwidth}
    \includegraphics[width=\textwidth]{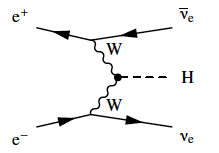}
    \caption{}
    \label{fig:hvv}
  \end{subfigure}
  \hfill
  \begin{subfigure}[h!]{0.38\textwidth}
    \includegraphics[width=\textwidth]{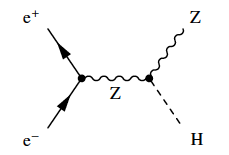}
    \caption{}
  \label{fig:hz}
  \end{subfigure}
  \caption{Feynman diagrams of the dominant Higgs production mechanisms above \subref{fig:hvv} and below \subref{fig:hz} 500 GeV center-of-mass energy.}\label{fig:4}
\end{figure}

Determination of the relative statistical uncertainty of the measurements $\mathrm{{\sigma(H\nu\bar{\nu})\times BR(H\rightarrow ZZ^\ast)}}$ at 350 GeV and 3 TeV CLIC is done for the semileptonic ${\mathrm{H\rightarrow ZZ^\ast}}$ final states, assuming a realistic luminosity spectrum and the presence of beamsstrahlung background. At 3 TeV, Higgs bosons are produced in WW-fusion (\cref{fig:hvv}) with a production cross-section of 415 fb. For 5 ab$^{-1}$ one thus expects around 6000 signal events. The signal signature at 3 TeV is qqll final state and missing energy. At 350 GeV Higgs boson is produced in Higgsstrahlung process (\cref{fig:hz}) with a cross section of about 93.44 fb. For 1 ab$^{-1}$ of available data one expects about 240 signal events with qqqqll final state where two jets are coming from a primary Z boson decaying hadronically.

\subsection{Event selection}

\begin{figure}
  \centering
  \begin{subfigure}[h!]{0.48\textwidth}
    \includegraphics[width=\textwidth]{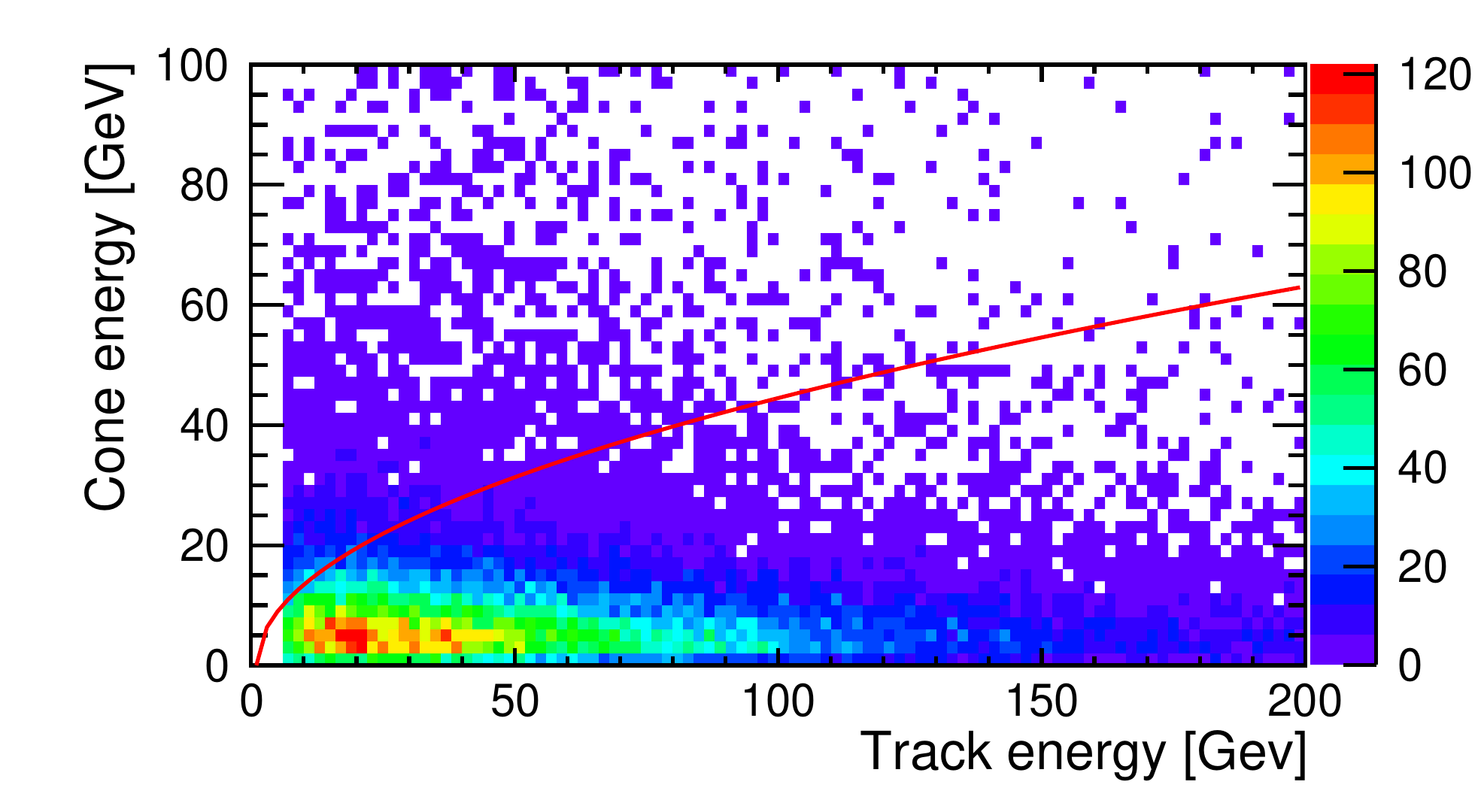}
        \begin{textblock}{4}(1., -2.3)
        \text{CLICdp, work in progress}
        \end{textblock}
    \caption{}
    \label{fig:c3t}
  \end{subfigure}
  \hfill
  \begin{subfigure}[h!]{0.48\textwidth}
    \includegraphics[width=\textwidth]{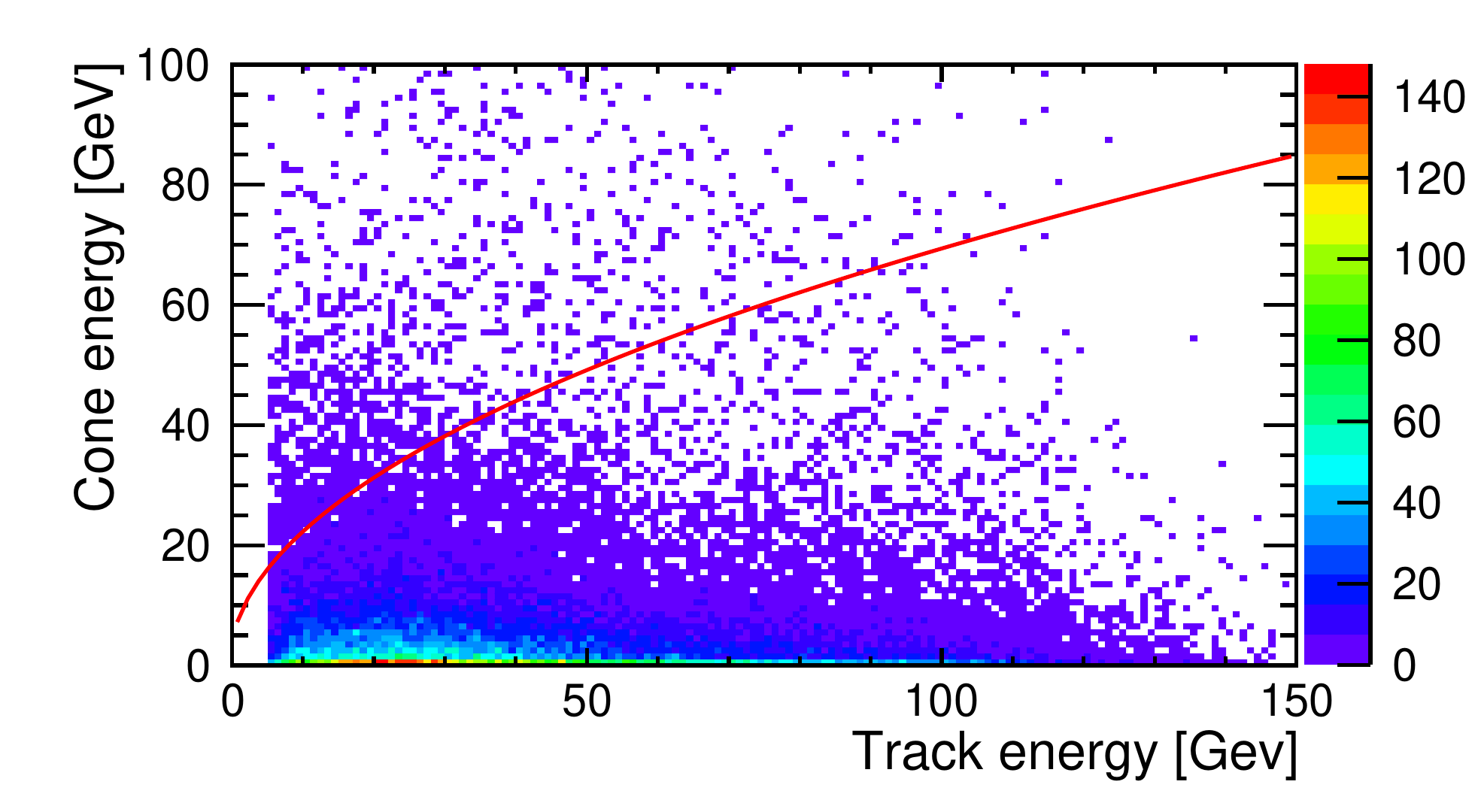}
        \begin{textblock}{4}(1., -2.3)
        \text{CLICdp, work in progress}
        \end{textblock}
    \caption{}
  \label{fig:c350g}
  \end{subfigure}
  \caption{Cone energy as a function of track energy of the reconstructed leptons at 3 TeV \subref{fig:c3t} and at 350 GeV center-of-mass energy \subref{fig:c350g}. The red line represents polynomial distribution separating isolated lepton tracks.}\label{fig:5}
\end{figure}

Since the Higgs decay products are the same in both measurements, event selections at 350 GeV and 3 TeV
share common methodolgy. Firstly, two leptons (electrons or muons) are isolated including recovery of
photons that are radiated by final state leptons (Bremsstrahlung recovery). Lepton dressing by adding
photons radiated in a 3\degrees cone improves the mass resolution of reconstructed on-shell Z bosons. The remaining particles are then grouped in 2 (4) jets by the k$_{\mathrm{T}}$ algorithm ~\cite{11}, where the cone radius of a jet is set to R = 0.7 (1.1) at 3 TeV (350 GeV) center-of-mass energy. In the preselection phase, we look for events with exactly
2 isolated leptons per event. Preselection primarily reduces backgrounds with large cross sections like $\gamma\gamma\rightarrow q\bar{q}$, $\gamma\gamma\rightarrow q\bar{q}l^+l^-$, at 3 TeV and $e^-e^+\rightarrow q\bar{q}q\bar{q}$, $e^-e^+\rightarrow q\bar{q}l^+l^-$, at 350 GeV. Lepton isolation is done with the Isolated Lepton Finder (ILF) Marlin processor ~\cite{12} that uses several
parameters in lepton isolation, including track energy of a particle (E$_{\mathrm{track}}$), ratio of energy deposited in
electromagnetic (ECAL) and hadronic (HCAL) calorimeters R$_{\mathrm{CAL}}$ ($\mathrm{R_{CAL} =E_{ECAL}/(E_{ECAL}+E_{HCAL})}$), longitudinal z$_\mathrm{0}$,
transverse d$_\mathrm{0}$, 3D impact parameter R$_\mathrm{0}$ ($\mathrm{R_0 = \sqrt{{z_0}^2+{d_0}^2}}$) and isolation curve removing from the preselection lepton candidates with too much energy in a cone around them (E$_\mathrm{cone}$ ). \cref{fig:5}
illustrates the distristibution of E$_\mathrm{cone}$ versus lepton track energy (E$_{\mathrm{track}}$). Since Beamsstrahlung background is
more pronounced at high energy, the isolation cone contains more energy at 3 TeV than at 350 GeV and
therefore loss of events with leptons on isolation curve is larger at the higher center-of-mass energy. This
effect is partially reduced with the additional requirement on minimal transverse momenta of particles in
the isolation cone.

\begin{figure}
  \centering
  \begin{subfigure}[h!]{0.48\textwidth}
    \includegraphics[width=1.5\linewidth]{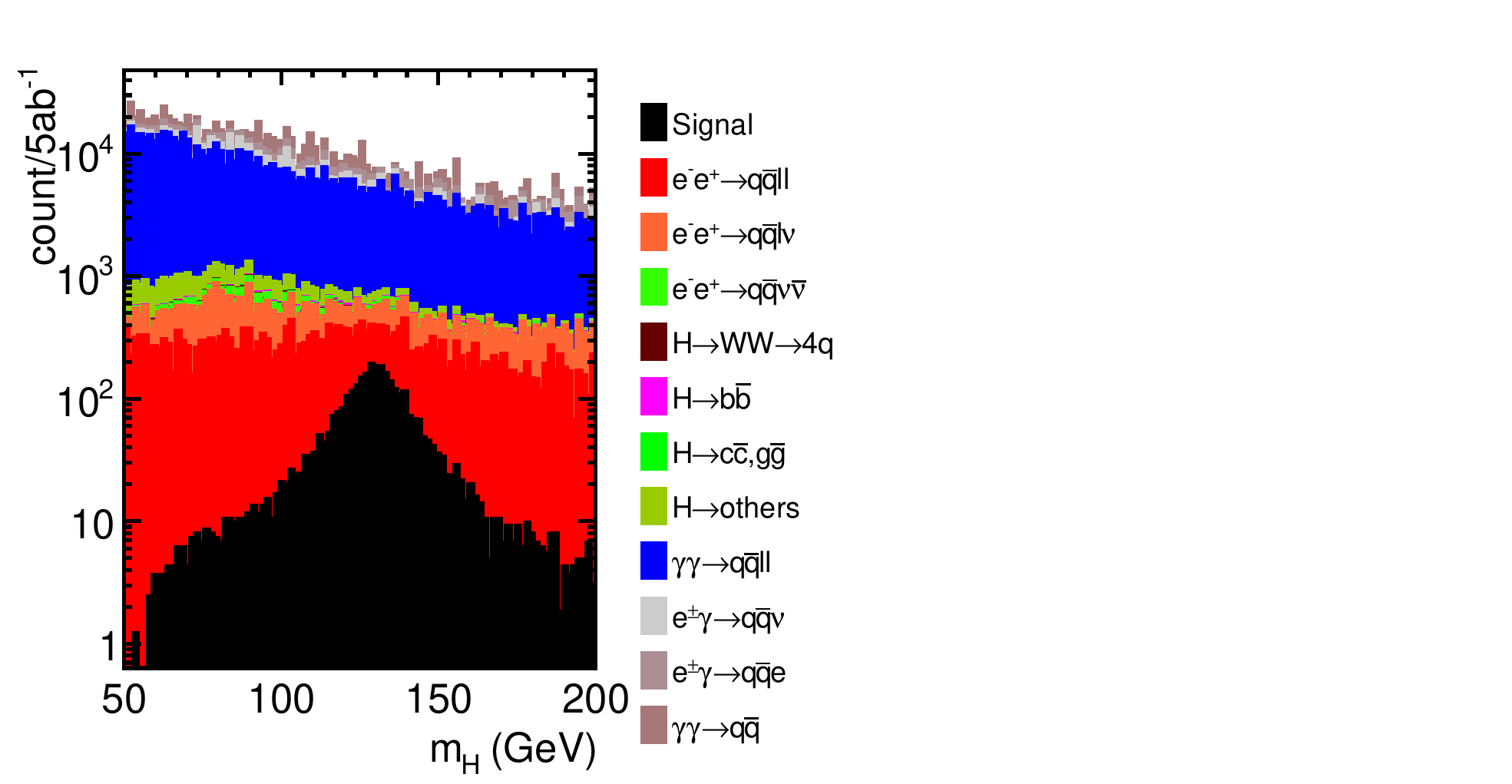}
        \begin{textblock}{4}(0.7, -3.2)
        \text{CLICdp, work in progress}
        \end{textblock}
    \caption{}
    \label{fig:pres3}
  \end{subfigure}
  \hfill
  \begin{subfigure}[h!]{0.48\textwidth}
    \includegraphics[width=1.5\linewidth]{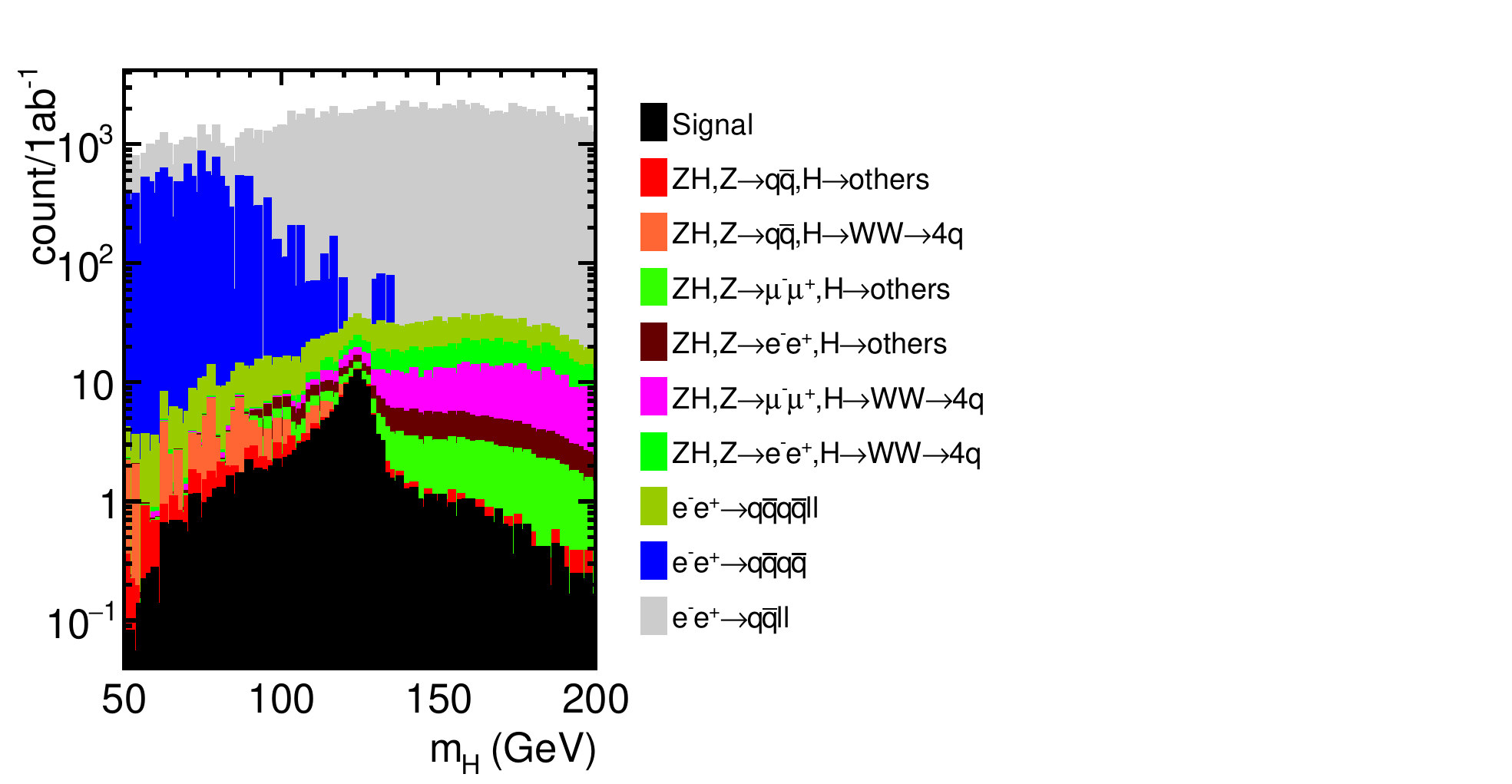}
        \begin{textblock}{4}(0.7, -3.2)
        \text{CLICdp, work in progress}
        \end{textblock}
    \caption{}
  \label{fig:pres350}
  \end{subfigure}
  \caption{Stacked histograms of the Higgs invariant mass distributions after preselection phase, at 3 TeV \subref{fig:pres3} and 350
  GeV \subref{fig:pres350}.}\label{fig:6}
\end{figure}

Preselection efficiencies are 67\% and 77\% at 3 TeV and 350 GeV, respectively. In \cref{fig:6} (a and b), stacked
histograms of the Higgs invariant mass are illustrated after the preselection phase at 3 TeV and 350 GeV,
respectively.

\subsection{Multivariate analysis}

\begin{figure}
  \centering
  \begin{subfigure}[!h]{0.48\textwidth}
    \includegraphics[width=1.5\linewidth]{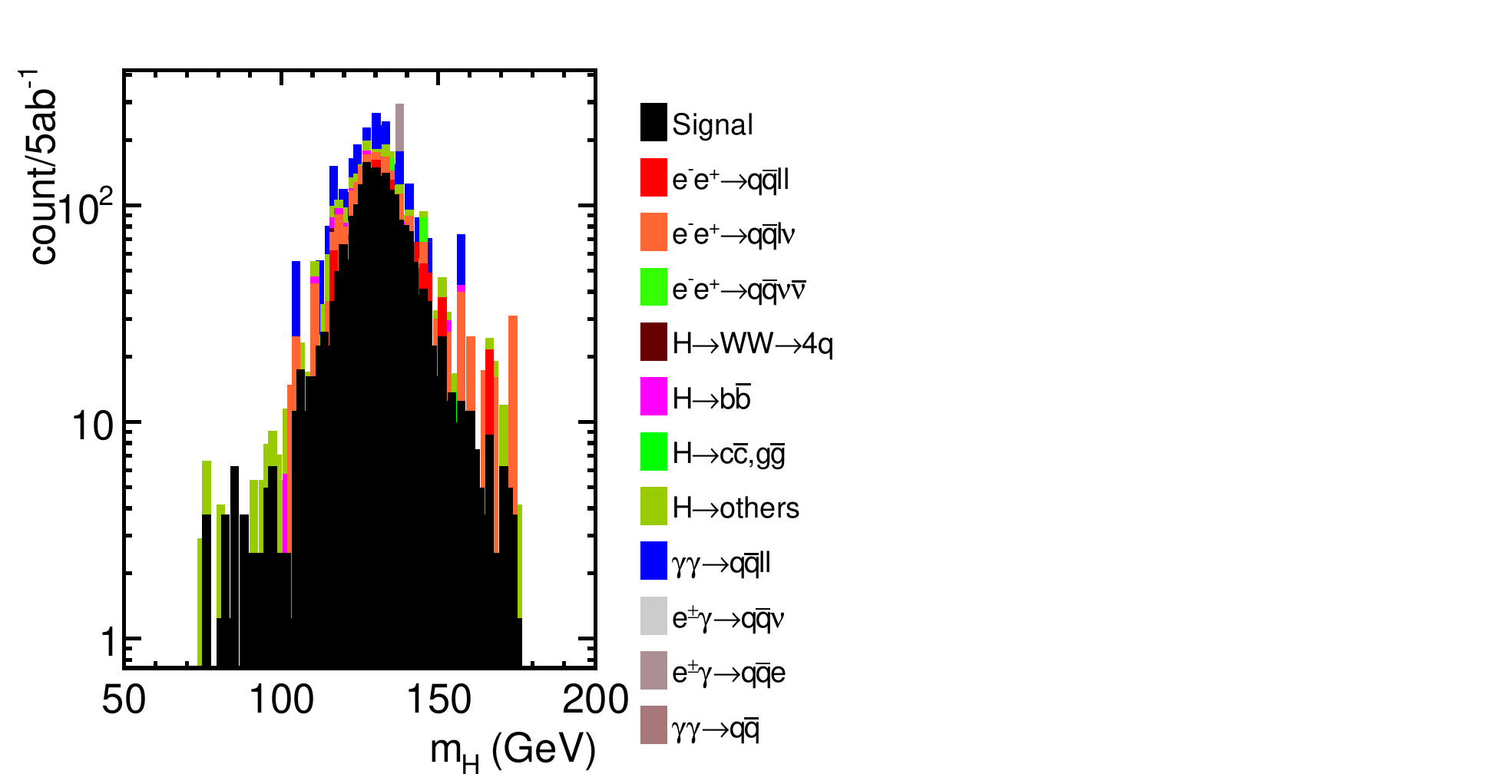}
        \begin{textblock}{3.5}(0.7, -3.2)
        \text{CLICdp, work in progress}
        \end{textblock}
    \caption{}
    \label{fig:mva3}
  \end{subfigure}
  \hfill
  \begin{subfigure}[!h]{0.48\textwidth}
    \includegraphics[width=1.5\linewidth]{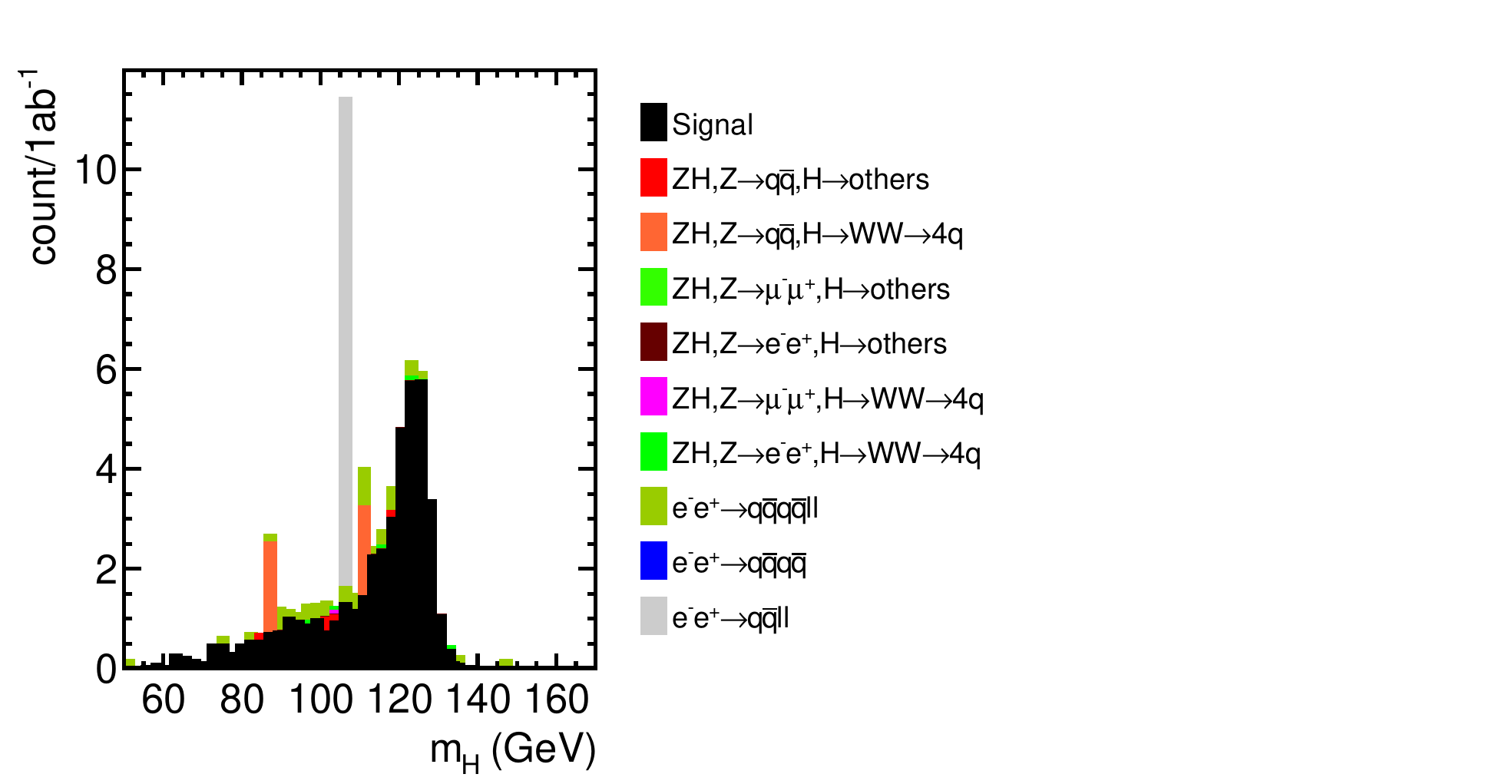}
        \begin{textblock}{3.5}(0.7, -3.2)
        \text{CLICdp, work in progress}
        \end{textblock}
    \caption{}
  \label{fig:mva350}
  \end{subfigure}
  \caption{Stacked histograms of the Higgs invariant mass distributions after MVA, at 3 TeV \subref{fig:mva3} and 350 GeV \subref{fig:mva350} center-of-mass energies.}\label{fig:7}
\end{figure}

Final separation of signal from background is done by employing a multivariate analysis (MVA). The Toolkit
for Multivariate Analysis (TMVA) ~\cite{13} is applied using the Boosted Decision Tree (BDT) method in
classification of events. It tends to maximize the statistical significance of signal to background separation.
The statistical uncertainty $\mathrm{\delta}$ of a measurement is derived from the statistical significance S as:

\begin{equation}
\label{delta}
\delta = \dfrac{1}{S} = \dfrac{\sqrt{N_{S}+N_{B}}}{N_{S}}
\end{equation}

\noindent where N$_{\mathrm{S(B)}}$ is the number of signal (background) events.
The BDT is trained on 16 (20) sensitive observables at 3 TeV (350 GeV), like masses of reconstructed Z bosons,
Higgs mass and polar angle, visible energy of an event, b and c-tagging probabilities of jets and jet
transition variables. At both energies, the Higgs mass is most sensitive to the signal-to-background separation,
that is done in a window around 126 GeV. In \cref{fig:7} (a and b), stacked histograms of the reconstructed
Higgs mass are given after MVA application, at 3 TeV (a) and 350 GeV (b) center-of-mass energies.

\begin{table}
\centering
\caption{Overview of the signal and the irreducible background process at 3 TeV and 350 GeV center-of-mass energies.}
\label{tab:pres_mva}
\begin{tabular}{l l l l }
\toprule
Process  & {$\epsilon_{presel}$} & {$\epsilon_{BDT}$} & {$N_{BDT}$} \\
\midrule
Signal $@$ 3 TeV & 67\% & 59\% & 2232 \\
\midrule
Background process \\
\midrule
$\gamma\gamma\rightarrow q\bar{q}l^+l^-$	& 11\textperthousand 	& 0.4\textperthousand	& 672    \\
$e^-e^+\rightarrow q\bar{q}l\nu$	& 3\textperthousand	& 6\textperthousand  & 509    \\
$e^-e^+\rightarrow H\nu\bar{\nu}; H\rightarrow others$	& 45\textperthousand	& 1.6\%	& 328  \\
$e^-e^+\rightarrow q\bar{q}l^+l^-$	& 7.5\textperthousand	& 1\textperthousand  & 126    \\
$e^\pm\gamma\rightarrow q\bar{q}e$	& 8.8\textperthousand	& 0.2\textperthousand	&  116	\\ 
processes with $N_{BDT}$<100	& 0.7\textperthousand	& 1.7\%	&  98   \\
\midrule
\\[1pt]
\midrule
Signal $@$ 350 GeV & 77\% & 23\% & 43 \\
\midrule
Background process \\
\midrule
$e^-e^+\rightarrow q\bar{q}l^+l^-$       & 11.4\%	& 0.05\textperthousand	& 10  \\
$e^-e^+\rightarrow q\bar{q}q\bar{q}l^+l^-$       & 21\%	& 5\textperthousand	& 5  \\
$e^-e^+\rightarrow HZ; Z\rightarrow q\bar{q}, H\rightarrow WW\rightarrow 4q$       & 0.42\%	& 9\%	& 4  \\
\bottomrule
\end{tabular}
\end{table}

\subsection{Statistical uncertainties}

Signal BDT efficiencies are estimated to be 59\% and 23\%, at 3 TeV and 350 GeV, respectively. The total signal
efficiency after all selection phases is 39\% at 3 TeV and 18\% at 350 GeV. The relatively low BDT efficiency at
350 GeV is due to the fact that the signal is very rare in nature (below 200 preselected events in 1 ab$^{-1}$), so the
relative statistical uncertainty is sensitive even to the smallest loss of signal. \cref{tab:pres_mva} gives
the expected number of selected signal and background events at 3 TeV and 350 GeV.
The relative statistical uncertainty derived from statistical significance as in Eq.(\ref{delta}), is found to be 3\% at 3 TeV
and 18\% at 350 GeV, assuming integrated luminosities of 5 ab$^{-1}$ and 1 ab$^{-1}$, respectively. With the proposed
polarization scheme of $\pm$80\% longitudinal electron-beam polarization and no positron polarization,
the relative statistical uncertainty of the 3 TeV measurement will be conservatively decreased by a factor
$\sqrt{1.48}$ ~\cite{14}, while the result at 350 GeV would not be relevantly influenced due to the different chiral nature of
the Higgs production mechanism.

\section{Conclusion}

The Higgs to ZZ$^\ast$ branching fraction measurement at CLIC is fully simulated at 350 GeV and 3 TeV center-of-
mass energies, for the semi-leptonic final states of Higgs to ZZ$^\ast$ decays. The relative statistical uncertainty of
a measurement is derived from the statistical significance and is found to be 3\% at 3 TeV and 18\% at 350
GeV, assuming integrated luminosities 5 ab$^{-1}$ and 1 ab$^{-1}$, respectively. The obtained result at 3 TeV is in line with the projection in ~\cite{15}.


\end{document}